\documentclass{PoS}
\rightline{\parbox{4cm}{\large\rm
ADP-13-22/T842\\
DESY 13-219\\
Edinburgh 2013/31\\
LTH 995
}}

\usepackage{amsmath}
\usepackage{amssymb}
\usepackage{graphicx}
\usepackage{wrapfig}
\newcommand{\be}{\begin{equation}}
\newcommand{\ee}{\end{equation}}
\newcommand{\bc}{\begin{center}}
\newcommand{\ec}{\end{center}}

\title{SU(3) flavour breaking and baryon structure}

\ShortTitle{SU(3) baryon structure}

\author{A.N.~Cooke$^{a}$, R.~Horsley$^{a}$, Y.~Nakamura$^{b}$,
  D.~Pleiter$^{c,d}$, P.~E.~L.~Rakow$^{e}$, P.~Shanahan$^{f}$,
  G.~Schierholz$^{g}$, H.~St\"uben$^{h}$,
  \speaker{J.~M.~Zanotti}$^{,f}$\\
        \llap{$^a$} School of Physics and Astronomy,
                    University of Edinburgh,
                    Edinburgh EH9 3JZ, UK \\
        \llap{$^b$} RIKEN Advanced Institute for Computational
                    Science, Kobe, Hyogo 650-0047, Japan \\
        \llap{$^c$} JSC, Forschungszentrum J\"ulich, 52425 J\"ulich,
                    Germany\\
        \llap{$^d$} Institut f\"ur Theoretische Physik,
                    Universit\"at Regensburg,
                    93040 Regensburg, Germany \\
        \llap{$^e$} Theoretical Physics Division,
                    Department of Mathematical Sciences,
                    University of Liverpool,
                    Liverpool L69 3BX, UK \\
        \llap{$^f$} CSSM, School of Chemistry and Physics, The
        University of Adelaide, Adelaide SA 5005, Australia\\
        \llap{$^g$} Deutsches Elektronen-Synchrotron DESY,
                    22603 Hamburg, Germany \\
        \llap{$^h$} Regionales Rechenzentrum, Universit\"at Hamburg,
                    20146 Hamburg, Germany \\
        E-mail: \email{james.zanotti@adelaide.edu.au}}

\author{QCDSF/UKQCD Collaboration}

\abstract{We present results from the QCDSF/UKQCD collaboration for
  hyperon electromagnetic form factors and axial charges obtained from
  simulations using $N_f=2+1$ flavours of $\mathcal{O}(a)$-improved
  Wilson fermions.
  We also consider matrix elements relevant for hyperon semileptonic
  decays.
  We find flavour-breaking effects in hyperon magnetic moments which
  are consistent with experiment, while our results for the connected
  quark spin content indicates that quarks contribute more to the spin
  of the $\Xi$ baryon than they do to the proton.  }

\FullConference{31st International Symposium on Lattice Field Theory -
  LATTICE 2013\\
  July 29 - August 3, 2013\\
  Mainz, Germany}

\begin{document}

\section{Introduction}

It has long been known that the nucleon is not a point-like object,
but in fact is composed of quarks and gluons.
However, many questions still remain, such as how are these
constituents distributed inside the nucleon? And how do they combine
to produce its experimentally observed properties?
For example, experimental results revealed that the quark spin
contribution to the spin of the proton is only about 30\%.  This was
originally termed the ``spin crisis'', however this quickly became
more of a ``spin puzzle'' as it was observed that quark orbital angular
momentum and gluon angular momentum would also contribute to the total
spin of the proton, although the exact decomposition remains unclear.
Given this observed suppression of the quark spin content of the
proton, an interesting quesion to ask is whether or not this is a
property unique to the nucleon, or a univeral feature of all hadrons.

Hence understanding how the nucleon and other hadrons are constructed
from their quark and gluon constituents remains one of the most
important and challenging questions in modern nuclear physics.
The study of the electromagnetic (EM) properties of hadrons, to cite
another example, provides important insights into the non-perturbative
structure of QCD.
The EM form factors reveal information on the internal structure of
hadrons including their size, charge distribution and magnetisation.
While only a few hyperon properties have been determined
experimentally (e.g. for the electric charge radii, only
$p,\,n,\,\Sigma^-$ have been measured), on the lattice they should (in
principle) be easier to calculate than nucleon observables due to the
presence of the heavier strange quark.

Lattice simulations have the potential to provide insights into the
charge and magnetic distribution of hyperons as well as the role of
SU(3)-flavour symmetry breaking in these distributions, which experiment
cannot currently provide.
These are of significant interest as they provide valuable
insights into the environmental sensitivity of the distribution of
quarks inside a hadron.
For example, how does the distribution of $u$ quarks in $\Sigma^+$
change as we change the mass of the (spectator) $s$ quark?
Simulations can also provide insights into the role of hidden flavour
(e.g. strangeness in the proton)
\cite{Leinweber:2004tc,Leinweber:2006ug}.
However, while the EM form factors of the nucleon have received a lot
of recent attention in lattice simulations (see, e.g.,
\cite{Hagler:2009ni} for a review), the investigation of the hyperon
EM form factors has so far received only limited attention
\cite{Boinepalli:2006xd,Lin:2008mr}.

Lattice simulations are currently performed in the isospin-symmetric
limit ($m_u=m_d$) (see \cite{Tantalo:2013maa} for recent progress in
isospin-breaking effects).
However, QCD is flavour-blind, so we could think of the $s$ quark as a
very heavy $d$ quark and compare lattice results for $\Sigma^+(uus)$
with the proton $(uud)$ to gain an idea of the implications for
isospin-breaking. 
This idea has already been explored in the context of the first
lattice determinations of charge-symmetry violation in moments of
spin-independent \cite{Horsley:2010th} and spin-dependent
\cite{Cloet:2012db} parton distribution functions.

Semileptonic form factors of the hyperons provide an alternative
method to the standard $K_{\ell 3}$ decays (see
e.g. \cite{Cabibbo:2003cu}) for determining the CKM matrix element,
$|V_{us}|$.
Earlier quenched and $N_f=2$ results for $\Sigma^-\to n\ell\nu$ and
$\Xi^0\to\Sigma^+\ell\nu$ can be found in
\cite{Guadagnoli:2006gj,Sasaki:2008ha}, while more recent $N_f=2+1$
results have been obtained in \cite{Sasaki:2012ne}.

In this talk we present preliminary results from the QCDSF/UKQCD
Collaboration for the octet hyperon electromagnetic and semileptonic
decay form factors, as well as their axial charges, determined from
$N_f=2+1$ lattice QCD. 

%
\section{Simulation Details}
\label{sec:simul}
%
Our gauge field configurations have been generated with $N_f=2+1$
flavours of dynamical fermions, using the tree-level Symanzik improved
gluon action and nonperturbatively ${\cal O}(a)$ improved Wilson
fermions \cite{Cundy:2009yy}.
We choose our quark masses by first finding the
$SU(3)_{\mathrm{flavour}}$-symmetric point where flavour singlet
quantities take on their physical values, then varying the individual
quark masses while keeping the singlet quark mass
$\overline{m}_q=(m_u+m_d+m_s)/3=(2m_l+m_s)/3$ constant
\cite{Bietenholz:2010jr}.
We have generated a large set of ensembles of varying quark masses and
volumes at $\beta=5.50$, corresponding to a lattice spacing,
$a=0.078(3)$~fm, where we have used the average baryon octet mass, $X_N$,
to set the scale \cite{Bietenholz:2011qq}.
We are now in the process of generating configurations at
additional lattice spacings to enable the continuum limit to be taken.
The results presented in this proceedings are obtained on a subset of
the complete set of ensembles.
%
%
A summary of the parameter space spanned by our dynamical
configurations can be found in Table~\ref{tab:results}.
More details regarding the tuning of our simulation parameters are
given in Refs.~\cite{Bietenholz:2010jr,Bietenholz:2011qq}.
\begin{table*}
\begin{center}
\begin{tabular}{c|c|c|c|c|c|c}
&$\kappa_l$ & $\kappa_s$ & $V=L^3\times T$ & $m_\pi$\,[MeV]& $m_K$\,[MeV] &
$m_\pi L$\\
\hline
1 & 0.120900 & 0.120900 & $32^3\times 64$ & 440 & 440 & 5.6 \\
2 & 0.121040 & 0.120620 & $32^3\times 64$ & 340 & 480 & 4.3 \\
3 & 0.121095 & 0.120512 & $32^3\times 64$ & 290 & 490 & 3.7 \\
4 & 0.121166 & 0.120371 & $48^3\times 96$ & 220 & 520 & 4.1
\end{tabular}
\caption{Simulation details for the subset of ensembles used here with
  $a=0.078(3)$\,fm \protect{\cite{Bietenholz:2011qq}}.}
\label{tab:results}
\end{center}
\end{table*}

In this preliminary work, we will not study systematic errors such as
potential contributions of disconnected diagrams, excited state
contamination or discretisation effects.
However, we hope that these effects are likely to be similar for
different members of the baryon octet and so perhaps are not
relevant for our investigation of SU(3)-breaking effects.

\section{Electromagnetic Form Factors}

On the lattice, we determine the form factors $F_1(q^2)$ and
$F_2(q^2)$ by calculating the following matrix element of the
electromagnetic current
\be
\langle B(p',\,s')| j_{\mu}(q)|B(p,\,s)\rangle
\, = 
 \bar{u}(p',\,s')
 \left[ \gamma_\mu F_1(q^2) +
       \sigma_{\mu\nu}\frac{q_\nu}{2M_B}F_2(q^2) \right] 
 u(p,\,s) \, ,
\label{eq:em-me}
\ee
where $u(p,\,s)$ is a Dirac spinor with momentum $p$, and spin
polarisation $s$, $q = p' - p$ is the momentum transfer, $M_B$ is the
mass of the baryon $B$, and $j_\mu$ is the electromagnetic current.
The Dirac $(F_1)$ and Pauli $(F_2)$ form factors of the proton are
obtained by using $j_\mu^{(p)} = \frac{2}{3}\bar{u}\gamma_\mu u -
\frac{1}{3}\bar{d}\gamma_\mu d$, while the form factors for the
$\Sigma$ and $\Xi$ baryons are obtained through the appropriate
substitution, $u\to s$ or $d\to s$.
It is common to rewrite the form factors $F_1$ and $F_2$ in terms of
the electric and magnetic Sachs form factors, 
$G_e= F_1 + q^2/(2M_N)^2\, F_2$ and $G_m= F_1 + F_2$.

If one is using a conserved current, then (e.g. for the proton)
$F_1^{(p)}(0) = G_e^{(p)}(0) =1$ gives the electric charge,
while $G_m^{(p)}(0) = \mu^{(p)} = 1 + \kappa^{(p)}$
gives the magnetic moment, where $F_2^{(p)}(0) = \kappa^{(p)}$ is the
anomalous magnetic moment.
From Eq.~(\ref{eq:em-me}) we see that $F_2$ always appears with a
factor of $q$, so it is not possible to extract a value for $F_2$ at
$q^2=0$ directly from our lattice simulations.
Hence we are required to extrapolate the results we obtain at finite
$q^2$ to $q^2=0$.

We perform dipole fits $F(q^2)=F(0)/(1-q^2/M^2)^2$ to $F_1$ and $F_2$,
although it was found in \cite{Collins:2011mk} that $F_1(q^2)$ and
$F_2(q^2)$ are better described by the ans\"atze $F_1(q^2) =
F_1(0)/(1+c_{12}q^2 + c_{14}q^4)$, and $F_2(q^2) =
F_2(0)/(1+c_{12}q^2 + c_{16}q^6)$.
This will be explored in future work.
%
%
%

\begin{figure}[t]
     \vspace*{-9mm}
     \hspace{-0.2mm}
   \begin{minipage}{0.47\textwidth}
     \hspace{-6mm}
          \includegraphics[clip=true,width=1.12\textwidth]{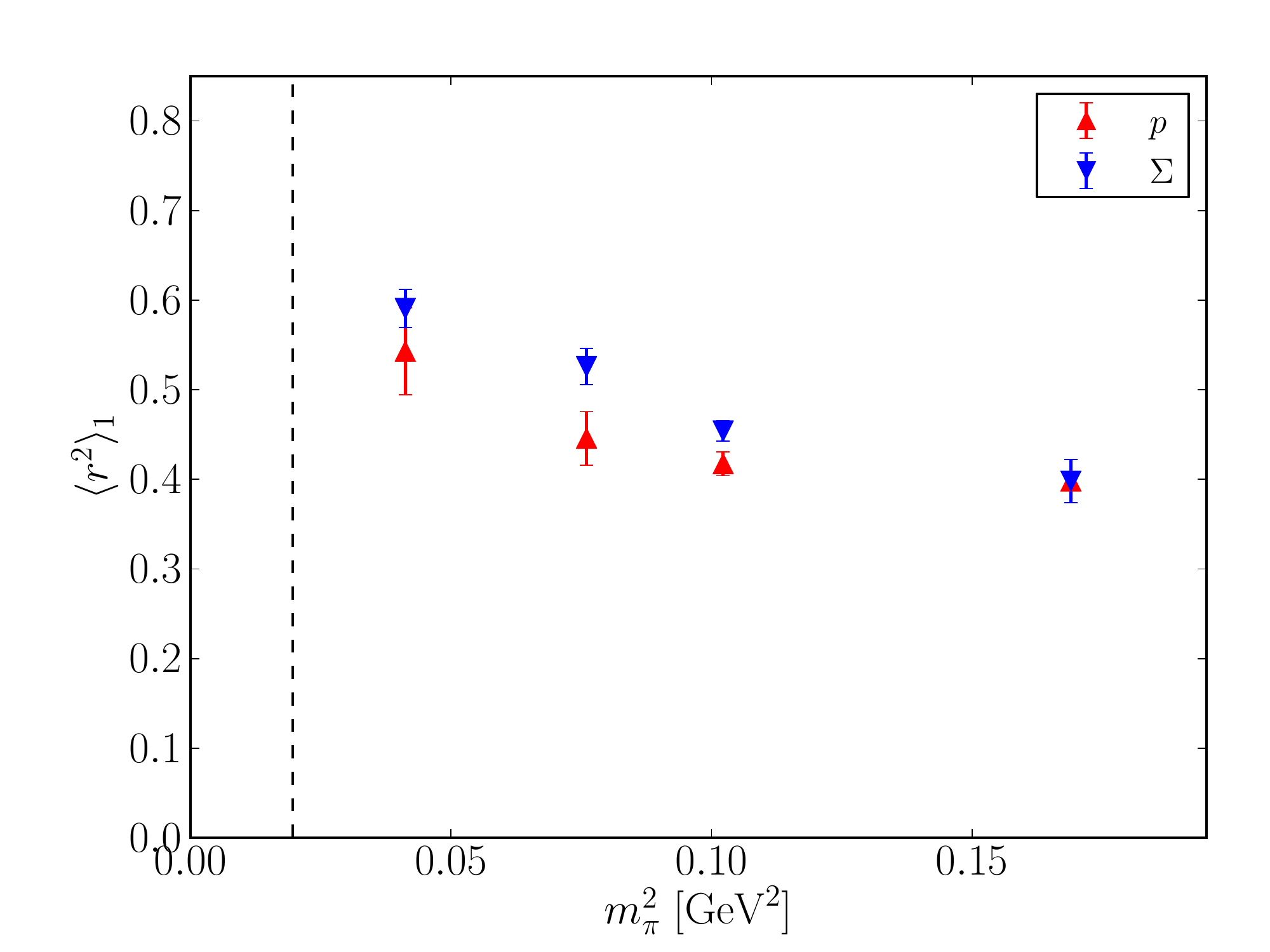}
          \caption{Dirac charge radii for the proton and
            $\Sigma^+$. Vertical dashed line represents the physical
            point.}
\label{fig:r1}
     \end{minipage}
     \hspace{3mm}
    \begin{minipage}{0.47\textwidth}
     \hspace{-5mm}
          \includegraphics[clip=true,width=1.12\textwidth]{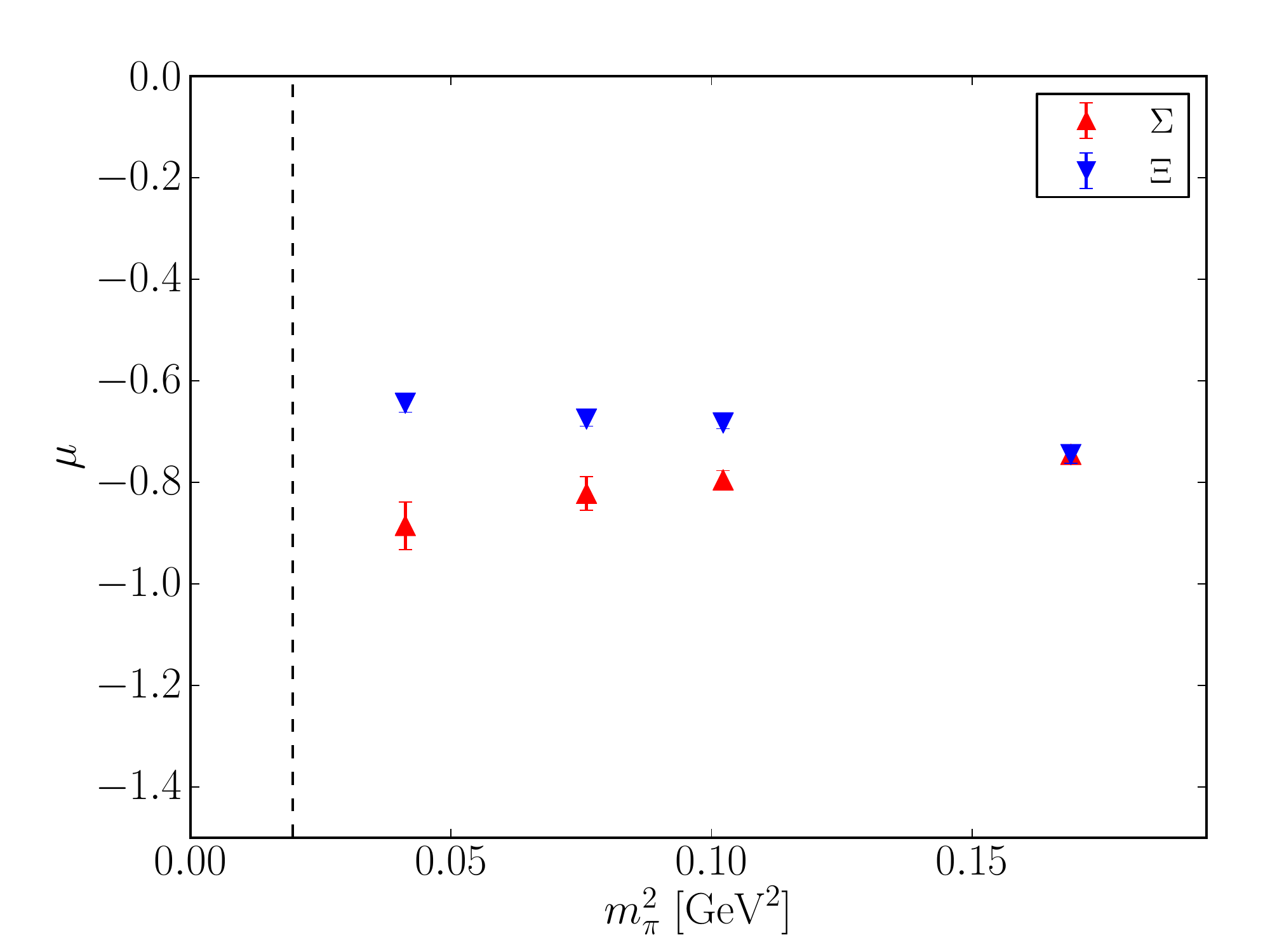}
          \caption{Magnetic moments for the $\Sigma^-$ and
            $\Xi^-$. Vertical dashed line represents the physical
            point.}
\label{fig:mm}
     \end{minipage}
 \end{figure}

%
Form factor radii, $r_i=\sqrt{\langle r^2\rangle_i}$, are defined from
the slope of the form factor at $q^2=0$.
In this talk, we are primarily interested in searching for any
$SU(3)$-flavour breaking effects in the octet hyperon form factors.
Fig.~\ref{fig:r1} shows the Dirac radii, $\langle r^2\rangle_1$,
of the proton and $\Sigma^+$.
This example is interesting as they are both charge $+1$, contain 2
up quarks, and one charge $-1/3$ quark. 
In the case of the proton this singly-represented quark is a light
down quark, while in the $\Sigma^+$ it is a heavier strange quark.
Starting from the SU(3)-symmetric limit where the $p$ and $\Sigma$ are
degenerate and hence have the same charge radius, we see similar quark
mass behaviour, but interestingly, we find the charge radius of the
$\Sigma^+$ to be slightly larger than that of the proton.

Fig.~\ref{fig:mm} shows the magnetic moments of the $\Sigma^-$ and
$\Xi^-$ hyperons.
These baryons are both charge $-1$, and contain three charge $-1/3$
quarks.
In the case of the $\Sigma^-$ this is 2 light and one strange, while
in the $\Xi^-$ this is reversed.
In the plot we see that the baryons have degenerate magnetic moments
in the SU(3)-flavour limit, but as we change the quark masses and
approach the physical point, denoted by the vertical dashed line, we
observe a definite splitting between the two magnetic moments, with
the ordering in agreement with experiment.

\section{Hyperon Axial Charges}

Hyperon axial charges are important for a low-energy effective field
theory description of octet baryons.
In the SU(3)-flavour limit, the hyperon axial charges are described by
two low energy constants, $F$ and $D$, e.g.
\begin{equation}
g_A^{NN}=F+D,\ g_A^{\Sigma\Sigma}=F,\ g_A^{\Xi\Xi}=F-D,\
g_A^{\Xi\Sigma}=F+D\ ,
\label{eq:F+D}
\end{equation}
where in our notation, the initial and final baryon states are given
as a superscript, so the standard nucleon axial charge is denoted,
$g_A^{NN}$, while the axial charge relevant to the $\Xi\to\Sigma$
semileptonic decay (see next section) is denoted, $g_A^{\Xi\Sigma}$.
The $F$ and $D$ constants, which describe all hyperon axial charges,
enter the chiral expansion of every baryonic quantity, e.g. masses.
Despite this, they are poorly determined.
Quark models, chiral perturbation theory, large-$N_c$ and fits to
hyperon beta decay give a range of results
\cite{Choi:2010ty,Close:1993mv,Gaillard:1984ny}, $F\approx 0.3-0.5,\
D\approx 0.6-0.8$.
Since we are interested in SU(3)-flavour breaking effects, we will be
looking for deviations from the expressions in Eq.~(\ref{eq:F+D}).
We have recently worked out the SU(3)-flavour breaking effects in
baryon matrix elements up to NLO \cite{Cooke:2012xv} which can be used
to fit lattice results once a sufficient collection of axial charges
is available.
%
%
Here we will investigate only a couple of axial charges.

\begin{figure}[t]
     \vspace*{-7mm}
     \hspace{-0.2mm}
   \begin{minipage}{0.47\textwidth}
     \hspace{-6mm}
          \includegraphics[clip=true,width=1.12\textwidth]{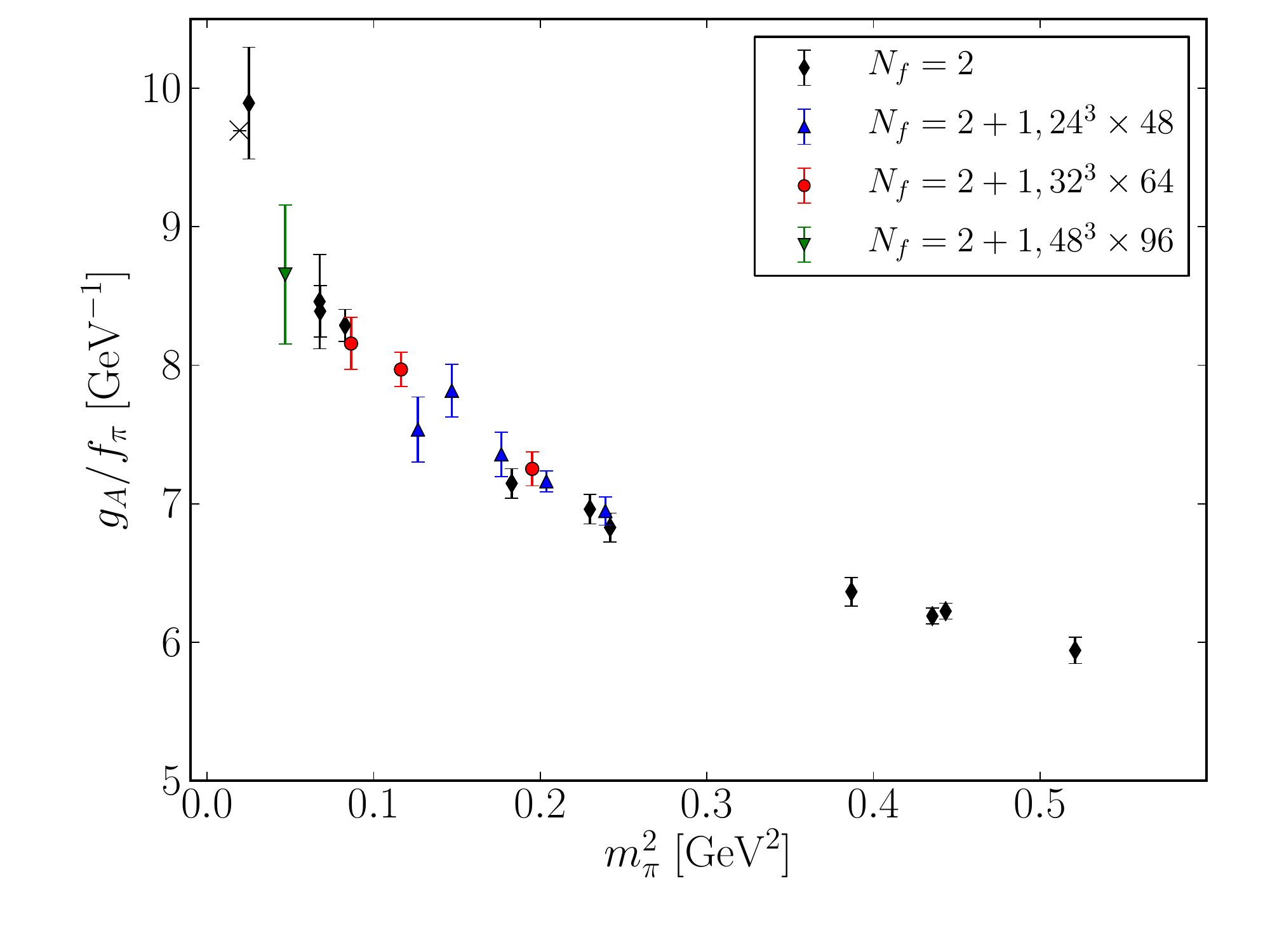}
          \caption{$N_f=2+1$ results for $g_A/f_\pi$ compared with
            $N_f=2$ results from \protect{\cite{Horsley:2013ayv}}.}
\label{fig:gAfpi}
     \end{minipage}
     \hspace{3mm}
    \begin{minipage}{0.47\textwidth}
     \vspace*{-4mm}
     \hspace{-5mm}
          \includegraphics[clip=true,width=1.12\textwidth]{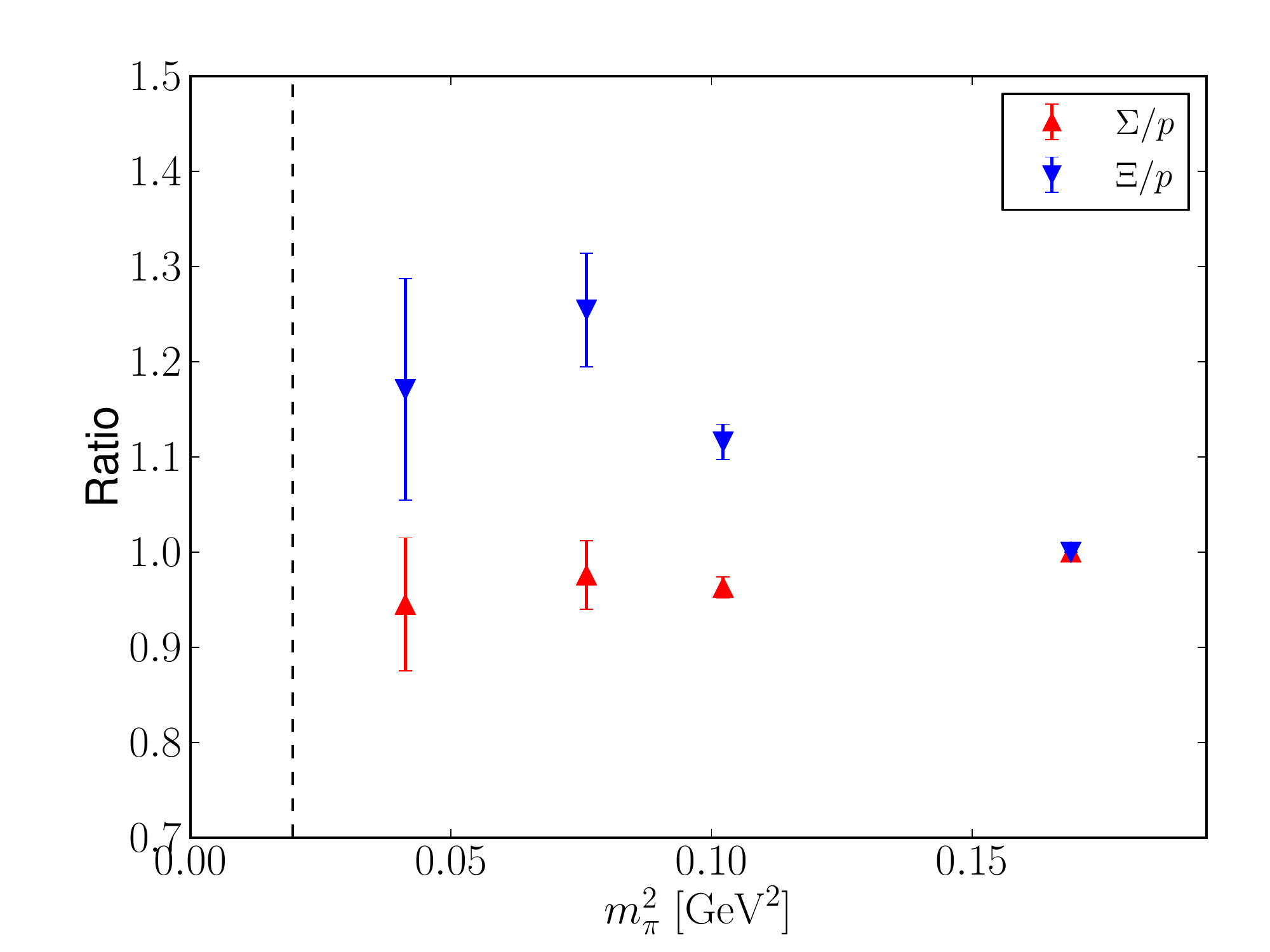}
          \caption{Ratio of the quark spin content of the $\Sigma$ and
          $\Xi$ to that of the proton.}
\label{fig:spin}
     \end{minipage}
 \end{figure}

The first quantity we will consider is the ratio $g_A/f_\pi$, as
proposed in \cite{Horsley:2013ayv}, as it is naturally renormalised and
has reduced finite size and discretisation effects.
In Fig.~\ref{fig:gAfpi} we compare our $N_f=2+1$ results from the
current work with the $N_f=2$ results from \cite{Horsley:2013ayv} and
we find excellent agreement, with a smooth trend towards the physical
point.

It is long known that intrinsic quark spin contributes only about 33\%
of the total spin of the proton.
In Fig.~\ref{fig:spin} we look for signs of similar spin suppression
in other hyperons by considering the ratio of the total connected
quark contribution to the $\Sigma$ and $\Xi$ spins to that of the
proton.
This idea was first proposed in \cite{Shanahan:2013apa} using some of
our earlier results from \cite{Cloet:2012db}.
Our new results exhibit similar findings, namely that the total
connected quark spin contribution in the $\Sigma$ is similar to that
of the proton, while a larger fraction of the $\Xi$ baryon spin comes
from the quark spin.
Hence this is a clear indication that quark spin suppression in
hadrons is not universal.

\section{Hyperon Semileptonic Form Factors}

Semileptonic form factors of the hyperons provide an alternative
method for determining the CKM matrix element, $|V_{us}|$.
This is done by using the experimental value for the decay rate of the
hyperon beta decays, $B\to b\ell\nu$, to obtain the combination
\begin{equation}
  \label{eq:vus}
  |V_{us}|^2|f_1(0)|^2\bigg( 1+3\left|\frac{g_1(0)}{f_1(0)}\right|^2
  \bigg)\ .
\end{equation}
Hence for a determination of $|V_{us}|$, we need to know the vector
and axial form factors, $f_1(q^2)$ and $g_1(q^2)$, at zero momentum
transfer $(q^2=0)$.

The vector and axial matrix elements for $SU(3)$-octet baryon
semileptonic decays are each given in terms of 3 form factors: vector
$(f_1)$, weak magnetism $(f_2)$, induced scalar $(f_3)$, axial-vector
$(g_1)$, weak electricity $(g_2)$ and induced pseudoscalar
$(g_3)$ (see e.g. \cite{Sasaki:2008ha}).
For a lattice calculation of hyperon beta decays, it is useful to
define the scalar form factor
%
$  f_0(q^2)=f_1(q^2) + \frac{q^2}{M_B^2 + M_b^2}f_3(q^2)$\ ,
%
which can be obtained at $q^2_{\mathrm{max}}=(M_B-M_b)^2$ with high
precision \cite{Hashimoto:1999yp}.
Interpolating lattice results obtained at finite $q^2$ values to
$q^2=0$ leads to the desired result $f_0(0)=f_1(0)$.
To reduce the systematic error involved in this $q^2$
interpolation, it has been shown that twisted boundary conditions
greatly help in the context of $K_{\ell 3}$ decays
\cite{Boyle:2010bh,Boyle:2013gsa}. 
For some of our ensembles, we have implemented this technique for
hyperon semileptonic decays in order to assess their effectiveness.

\begin{figure}[t]
     \vspace*{-9mm}
     \hspace{-0.2mm}
   \begin{minipage}{0.47\textwidth}
     \hspace{-6mm}
          \includegraphics[clip=true,width=1.12\textwidth]{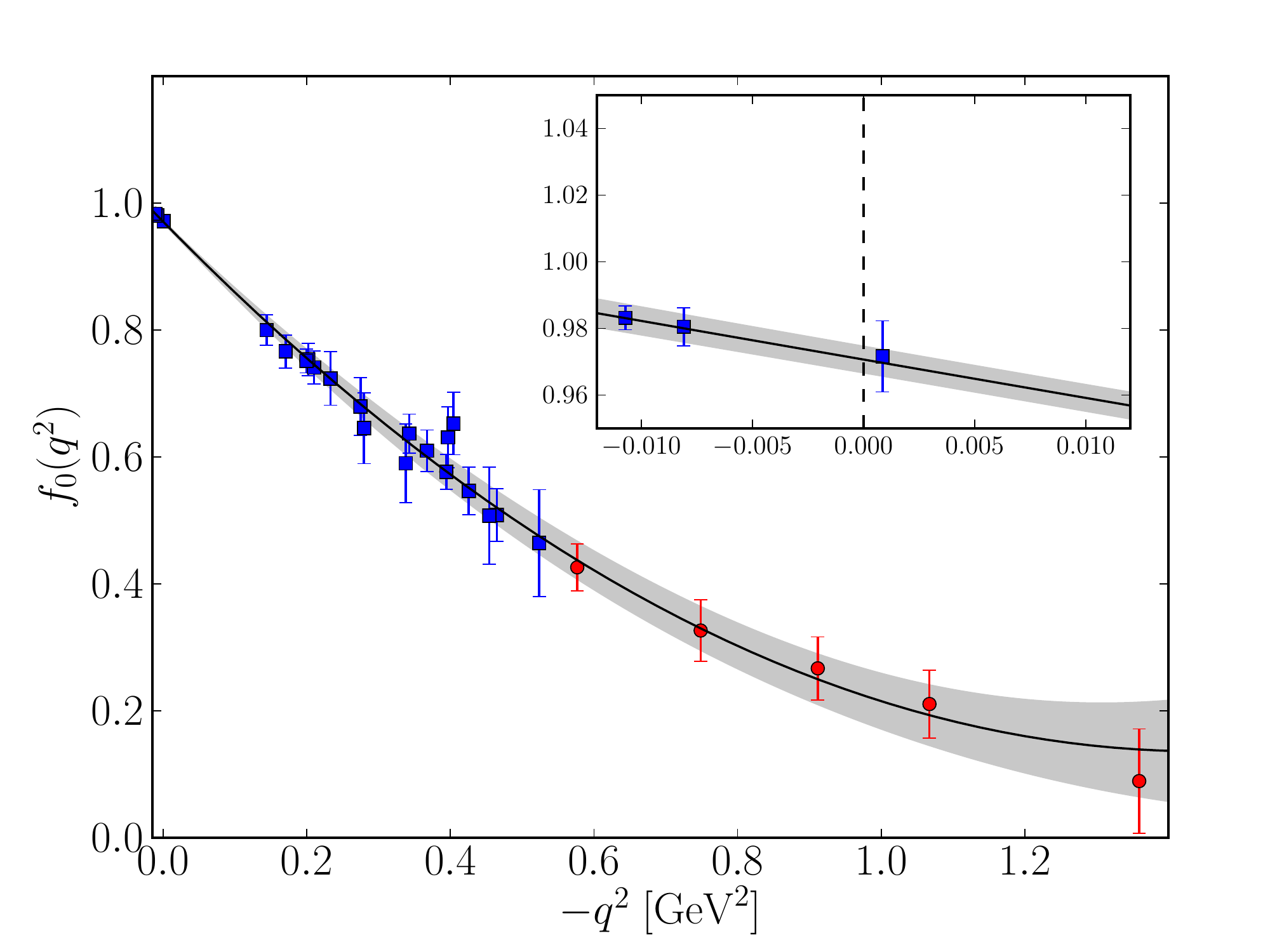}
          \caption{$f_0(q^2)$ for the $\Sigma^-\to n\ell\nu_\ell$
            decay from a $32^3\times 64$ lattice with $m_\pi\approx
            340$~MeV.}
\label{fig:f0q2}
     \end{minipage}
     \hspace{3mm}
    \begin{minipage}{0.47\textwidth}
     \hspace{-5mm}
          \includegraphics[clip=true,width=1.12\textwidth]{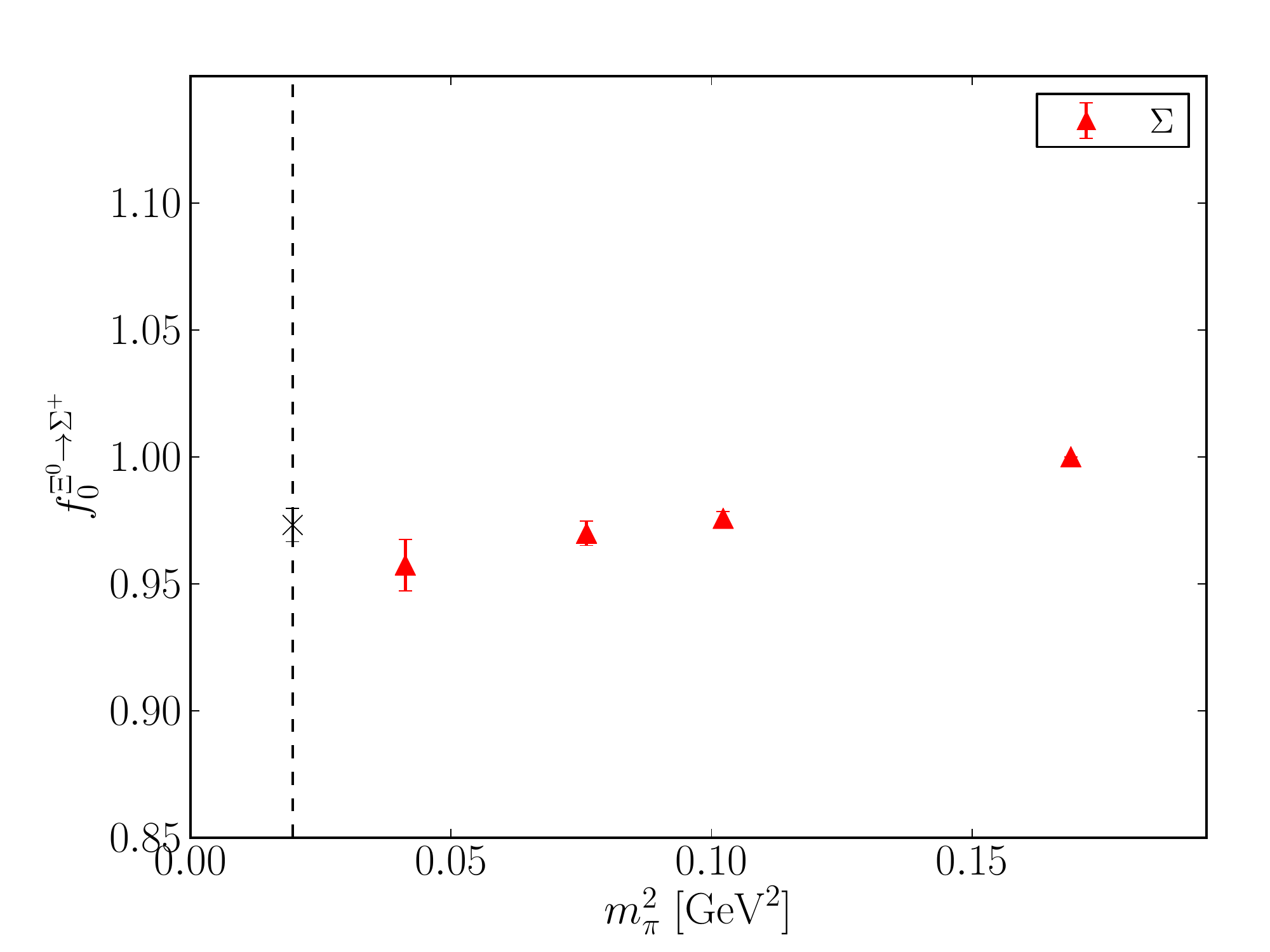}
          \caption{Quark mass dependence of $f_0(0)=f_1(0)$ for the
            $\Xi^0\to\Sigma^+\ell\nu_\ell$ decay.}
\label{fig:f0qm}
     \end{minipage}
 \end{figure}

In Fig.~\ref{fig:f0q2} we present some results for $f_0(q^2)$ for the
$\Sigma^-\to n\ell\nu_\ell$ decay from a $32^3\times 64$ lattice with
$m_\pi\approx 340$~MeV.
The red circles indicate results obtained using only Fourier
momenta, while results obtained using twisted boundary conditions are
given by blue squares.
We zoom into the region of interest around $q^2=0$ in the insert, where
we see that our choice of twists enable us to obtain a result directly
at $q^2=0$.
%

After interpolating $f_0(q^2)$ to $q^2=0$, we investigate the quark
mass dependence of $f_0(0)$ in Fig.~\ref{fig:f0qm} and compare our
result to the extrapolated result from \cite{Sasaki:2012ne}.
Although we haven't yet performed our chiral extrapolation, we will
complete this soon in the context of SU(3)-flavour breaking expansions
\cite{Cooke:2012xv}.

\section*{Acknowledgements}

The numerical configuration generation was performed using the BQCD
lattice QCD program, \cite{Nakamura:2010qh}, on the IBM BlueGeneQ using
DIRAC 2 resources (EPCC, Edinburgh, UK), the BlueGene P and Q at NIC
(J\"ulich, Germany) and the SGI ICE 8200 at HLRN (Berlin--Hannover,
Germany). The BlueGene codes were optimised using Bagel \cite{Bagel}.
The Chroma software library \cite{Edwards:2004sx}, was used in the
data analysis.
This investigation has been supported partly by the EU grants 283286
(HadronPhysics3), 227431 (Hadron Physics2) and 238353 (ITN
STRONGnet).
JMZ is supported by the Australian Research Council grant
FT100100005. We thank all funding agencies.


\begin{thebibliography}{99}

\bibitem{Leinweber:2004tc}
  D.~B.~Leinweber  {\it et al.},
  Phys.\ Rev.\ Lett.\  {\bf 94} (2005) 212001
  [hep-lat/0406002].

\bibitem{Leinweber:2006ug}
  D.~B.~Leinweber  {\it et al.},
  Phys.\ Rev.\ Lett.\  {\bf 97} (2006) 022001
  [hep-lat/0601025].

\bibitem{Hagler:2009ni}
  Ph.~H\"agler,
  Phys.\ Rept.\  {\bf 490 } (2010)  49
  [0912.5483 [hep-lat]].

\bibitem{Boinepalli:2006xd}
  S.~Boinepalli {\it et al.},
  Phys.\ Rev.\  {\bf D74 } (2006)  093005
  [hep-lat/0604022].

\bibitem{Lin:2008mr}
  H.~-W.~Lin and K.~Orginos,
  Phys.\ Rev.\  {\bf D79 } (2009)  074507
  [0812.4456 [hep-lat]].

\bibitem{Tantalo:2013maa}
  N.~Tantalo,
  PoS LATTICE {\bf 2013} (2013) 007  
  [1311.2797 [hep-lat]].

\bibitem{Horsley:2010th}
  R.~Horsley {\it et al.},
  Phys.\ Rev.\ D {\bf 83} (2011) 051501
  [1012.0215 [hep-lat]].

\bibitem{Cloet:2012db}
  I.~C.~Cloet {\it et al.},
  Phys.\ Lett.\ B {\bf 714} (2012) 97
  [1204.3492 [hep-lat]].

\bibitem{Cabibbo:2003cu}
  N.~Cabibbo, E.~C.~Swallow and R.~Winston,
  Ann.\ Rev.\ Nucl.\ Part.\ Sci.\  {\bf 53} (2003) 39
  [hep-ph/0307298].

\bibitem{Guadagnoli:2006gj}
  D.~Guadagnoli {\it et al.}, 
  Nucl.\ Phys.\  B {\bf 761} (2007) 63
  [hep-ph/0606181].

\bibitem{Sasaki:2008ha}
  S.~Sasaki and T.~Yamazaki,
  Phys.\ Rev.\  {\bf D79 } (2009)  074508
  [0811.1406 [hep-ph]].

\bibitem{Sasaki:2012ne}
  S.~Sasaki,
  Phys.\ Rev.\ D {\bf 86} (2012) 114502
  [1209.6115 [hep-lat]].

\bibitem{Cundy:2009yy}
  N.~Cundy {\it et al.} [QCDSF/UKQCD Collaboration],
  Phys.\ Rev.\  {\bf D79 } (2009)  094507
  [0901.3302 [hep-lat]].

\bibitem{Bietenholz:2010jr}
  W.~Bietenholz {\it et al.} [QCDSF/UKQCD Collaboration],
  Phys.\ Lett.\  {\bf B690 } (2010)  436
  [1003.1114 [hep-lat]].

\bibitem{Bietenholz:2011qq}
  W.~Bietenholz {\it et al.} [QCDSF/UKQCD Collaboration],
  Phys.\ Rev.\ D {\bf 84} (2011) 054509
  [1102.5300 [hep-lat]].

\bibitem{Collins:2011mk}
  S.~Collins {\it et al.} [QCDSF Collaboration],
  Phys.\ Rev.\ D {\bf 84} (2011) 074507
  [1106.3580 [hep-lat]].

\bibitem{Choi:2010ty}
  K.~-S.~Choi, W.~Plessas and R.~F.~Wagenbrunn,
  Phys.\ Rev.\ D {\bf 82} (2010) 014007
  [1005.0337 [hep-ph]].

\bibitem{Close:1993mv}
  F.~E.~Close and R.~G.~Roberts,
  Phys.\ Lett.\ B {\bf 316} (1993) 165
  [hep-ph/9306289].

\bibitem{Gaillard:1984ny}
  J.~M.~Gaillard and G.~Sauvage,
  Ann.\ Rev.\ Nucl.\ Part.\ Sci.\  {\bf 34 } (1984)  351.

\bibitem{Cooke:2012xv}
  A.~N.~Cooke {\it et al.} [QCDSF Collaboration],
  PoS LATTICE {\bf 2012} (2012) 116
  [1212.2564 [hep-lat]].


\bibitem{Horsley:2013ayv}
  R.~Horsley {\it et al.} [QCDSF/UKQCD Collaboration],
  1302.2233 [hep-lat].

\bibitem{Shanahan:2013apa}
  P.~E.~Shanahan {\it et al.},
  Phys.\ Rev.\ Lett.\  {\bf 110} (2013) 202001
  [1302.6300 [nucl-th]].

\bibitem{Hashimoto:1999yp}
  S.~Hashimoto {\it et al.},
  Phys.\ Rev.\  {\bf D61 } (1999)  014502
  [hep-ph/9906376].

\bibitem{Boyle:2010bh}
  P.~A.~Boyle {\it et al.}  [RBC/UKQCD Collaboration],
  Eur.\ Phys.\ J.\ C {\bf 69} (2010) 159
  [1004.0886 [hep-lat]].

\bibitem{Boyle:2013gsa}
  P.~A.~Boyle {\it et al.} [RBC/UKQCD Collaboration],
  JHEP {\bf 1308} (2013) 132
  [1305.7217 [hep-lat]].


\bibitem{Nakamura:2010qh}
  Y.~Nakamura and H.~St\"uben,
  PoS LATTICE {\bf 2010} (2010) 040
  [1011.0199 [hep-lat]].

\bibitem{Bagel}
  P.~A.~Boyle, 
  Comp.\ Phys.\ Comm. {\bf 180}, (2009) 2739.

\bibitem{Edwards:2004sx}
  R.~G.~Edwards and B.~Jo\'o,
  Nucl.\ Phys.\ Proc.\ Suppl.\  {\bf 140 } (2005)  832
  [hep-lat/0409003].

\end{thebibliography}
\end{document}